\documentclass[prl,superscriptaddress,twocolumn]{revtex4}
\usepackage{enumerate}
\usepackage{amsfonts,amssymb,amsmath}
\usepackage[]{graphics,graphicx,epsfig}
\usepackage{amsthm}
\newtheorem{theorem}{Theorem}

\usepackage{graphicx}
\usepackage{dcolumn}
\usepackage{natbib}
\usepackage{color}
\usepackage{multirow}
\usepackage{ulem}

\begin{document}


\title{Complementarity in quantum walks}

\author{Andrzej Grudka}
\affiliation{Institute of Spintronics and Quantum Information, Faculty of Physics, Adam Mickiewicz University, 61-614 Pozna\'n, Poland}

\author{Pawe{\l} Kurzy{\'n}ski}
\affiliation{Institute of Spintronics and Quantum Information, Faculty of Physics, Adam Mickiewicz University, 61-614 Pozna\'n, Poland}

\author{Tomasz P. Polak}
\affiliation{Institute of Spintronics and Quantum Information, Faculty of Physics, Adam Mickiewicz University, 61-614 Pozna\'n, Poland}

\author{Adam S. Sajna}
\affiliation{Department of Theoretical Physics, Faculty of Fundamental Problems of Technology, Wroc{\l}aw University of Science and Technology, 50-370 Wroc{\l}aw, Poland}

\author{Jan W{\'o}jcik}
\affiliation{Faculty of Physics, Adam Mickiewicz University, 61-614 Pozna\'n, Poland}

\author{Antoni W{\'o}jcik}
\affiliation{Institute of Spintronics and Quantum Information, Faculty of Physics, Adam Mickiewicz University, 61-614 Pozna\'n, Poland}

\date{\today}


\begin{abstract}
We study discrete-time quantum walks on $d$-cycles with a position and coin-dependent phase-shift. Such a model simulates a dynamics of a quantum particle moving on a ring with an artificial gauge field. In our case the amplitude of the phase-shift is governed by a single discrete parameter $q$. We solve the model analytically and  observe that for prime $d$ there exists a strong complementarity property between the eigenvectors of two quantum walk evolution operators that act in the $2d$-dimensional Hilbert space. Namely, if $d$ is prime the corresponding eigenvectors of the evolution operators obey $|\langle v_q|v'_{q'} \rangle| \leq 1/\sqrt{d}$ for $q\neq q'$ and for all $|v_q\rangle$ and $|v'_{q'}\rangle$. We also discuss dynamical consequences of this complementarity. Finally, we show that the complementarity is still present in the continuous version of this model, which corresponds to a one-dimensional Dirac particle.
\end{abstract}

\maketitle



\section*{Introduction}

Quantum walks (QWs) are dynamical models that describe quantum particles moving on a lattice. In this work we focus on their discrete-time versions (DTQWs) \cite{Aharonov:1993td,Meyer_1996}. Such models can simulate dynamics of various physical systems, e.g. \cite{PhysRevA.103.012201,Xiao:2017te,PhysRevA.82.033429,Crespi:2013vv,PhysRevA.78.022314,Zhang_2016,PhysRevLett.93.180601,Arnault_2020,J. Math. Phys. C,PhysRevLett.111.160601}, and are known to be capable of universal quantum computation \cite{PhysRevLett.102.180501,PhysRevA.81.042330,doi:10.1126/science.1229957,Singh:2021ud}, and as a consequence, of universal quantum simulation \cite{doi:10.1126/science.1177838}. Moreover, they were already implemented on many experimental platforms \cite{Manouchehri_2014}. One of the most appealing features of DTQWs is the fact that relatively simple and finite models can be used to investigate highly-nontrivial and complex phenomena. In this sense DTQWs are considered to be quantum analogues of classical cellular automata \cite{Huerta-Alderete:2020va}

We study a DTQW on a $d$-cycle in which a single particle acquires a phase-shift that depends on its position and on a state of its coin. The coin is an auxiliary degree of freedom that decides wether the particle moves right or left. It is described by a two-dimensional subsystem, therefore the DTQW's Hilbert space has dimension $D=2d$. We show that despite an apparent simplicity, the model exhibits complex properties. Namely, we find that eigenvectors of the evolution operator exhibit a peculiar dependence on the amplitude of the phase-shift $\varphi = 2\pi q/d$, where $q=0,1,\ldots d-1$ is a parameter that governs the amplitude. In particular, we observe that for prime $d$ the two different DTQW evolution operators, governed by $q \neq q'$, the corresponding eigenvectors obey a strong complementarity property, i.e., $|\langle v_q|v'_{q'} \rangle| \leq 1/\sqrt{d}$ for all $|v_q\rangle$ and $|v'_{q'}\rangle$. 
 
Complementarity is a key feature of quantum theory that is fundamentally related to uncertainty. Strong complementarity between eigenvectors of two observables implies strong restrictions on their joint measurability. If the two $D$-dimensional observables are maximally complementary, their eigenvectors form Mutually Unbiassed Bases (MUB) \cite{DURT:2010wm}, i.e., the modulus of the overlap between any vector from one basis with any vector from the other basis is equal to $1/\sqrt{D}$. Complementary observables play an important role in quantum information science. Primarily, they are the cornerstone of quantum cryptography \cite{RevModPhys.74.145,Pirandola:20}. In addition, the Quantum Fourier Transform \cite{Nielsen:2010tx}, which changes between the eigenbases of two such observables, is the main ingredient of the Shor's factoring algorithm \cite{Shor:1994tb}. However, there are still many questions to be answered about the fundamental properties of complementarity. For example, it is still unknown how many MUB can be found in Hilbert spaces of dimension that is not a power of a prime \cite{DURT:2010wm}. Moreover, the physical interpretation of complementary observables is problematic. Besides position and momentum, or spin-$1/2$ projections onto three mutually orthogonal axes, strongly complementary observables do not have any intuitive physical interpretation \cite{Kurzynski:2010vq}. Our results shed some light on the latter problem. We show that strong complementarity can exist between evolution operators and that it can be induced by relatively small alterations of these operators. Moreover, the eigenvectors of the original and the altered evolution operators form almost Mutually Unbiassed Bases (aMUB), since they obey  $|\langle v_q|v'_{q'} \rangle| \leq \sqrt{2/D}$.

Although DTQWs can be implemented in laboratories, as far as we know they do not occur naturally in the universe. Therefore, in the second part of this work we focus on the DTQW continuous limit, which is known to describe a Dirac particle \cite{KURZYNSKI20086125,PhysRevA.73.054302} -- an elementary quantum relativistic system such as electron. We find that the observed complementarity relations still occur in the continuous version for which the factorability of $d$ is not an issue anymore. More precisely, we show that eigenvectors of Hamiltonians of one-dimensional Dirac particles subjected to different gauge fields obey analogous complementarity relations as in DTQW case for prime $d$. 


\section*{Description of the model}

We consider a one-dimensional DTQW on a $d$-cycle \cite{Venegas-Andraca:2012up}. The system consists of a particle that can be located at one of $d$ positions $x=0,1,\ldots,d-1$ (we assume $x \equiv x ~\text{mod}~ d$) and of a coin, a two-level system that can be in one of two states $b=\pm 1$. We represent these states in the following way: $|+\rangle = (1~0)^T$ and $|-\rangle = (0~1)^T$. The coin can be either a particle's internal degree of freedom, akin to a spin, or an external system. However, this choice is of no importance here. The general pure state of the system at time $t$ is given by
\begin{equation}
|\psi_t\rangle = \sum_{x=0}^{d-1}\sum_{b=\pm 1} \alpha_{x,b}(t) |x\rangle \otimes |b\rangle.
\end{equation}

A single step of the evolution is generated by the unitary operator
\begin{equation}
|\psi_{t+1}\rangle = U |\psi_{t}\rangle. 
\end{equation}
It is of the form
\begin{equation}\label{U}
U = S (\openone \otimes C) F,
\end{equation}
where 
\begin{equation}
S = \sum_x \left(|x+1\rangle\langle x| \otimes \begin{pmatrix}
1 & 0 \\ 0 & 0 \end{pmatrix} + |x-1\rangle\langle x| \otimes \begin{pmatrix}
0 & 0 \\ 0 & 1 \end{pmatrix} \right)
\end{equation}
is the conditional translation and $C$ is an arbitrary quantum coin toss operator 
\begin{equation}\label{coin}
C = e^{i\delta} \begin{pmatrix}c & s\\-s^{*} & c^{*}\end{pmatrix}\in \mathcal{U}(2),
\end{equation}
where
\begin{equation}
    c=\cos\theta e^{i\gamma},~s=\sin\theta e^{i\sigma},
\end{equation}
and $\delta$, $\theta$, $\gamma$, $\sigma$ are arbitrary angles. Finally, $F$ is a position and coin-dependent phase-shift operator
\begin{equation}
F = \sum_x |x\rangle\langle x| \otimes \begin{pmatrix}
e^{i\varphi x} & 0 \\ 0 & e^{-i\varphi x} \end{pmatrix},~~~~\left(\varphi = \frac{2\pi q}{d}= \varepsilon q \right).
\end{equation}
The parameter $q=0,1,\dots,d-1$ determines the phase-shift's magnitude.

The above discrete spacetime model simulates a particle hopping on a ring that is subjected to an artificial gauge field. The gauge field is generated by $F$ and its magnitude is given by $\varphi = \varepsilon q$. We will come back to this interpretation later when we will discuss the continuous spacetime limit.  


\section*{Results}

The eigen-problem for the unitary evolution operator (\ref{U}) is studied in the Appendix A. Here we present its solution for d-prime and $\theta\neq 0$. The eigenvalues for a given $q$ are
\begin{equation}
\lambda_{m,\tau}^{(q)} = e^{i\left(\delta+q\frac{\varepsilon}{2} +\tau \xi_m^{(q)}\right)},
\end{equation}
where $\tau =\pm1$, $m=0,...,d-1$ and
\begin{equation}
    \cos \xi_{m}^{(q)}=(-1)^q\cos\theta\cos\left((m+q)\varepsilon-\gamma\right)
\end{equation}
The corresponding eigenvectors are
\begin{equation}\label{ev1}
|\psi_{m,\tau}^{(q\neq 0)}\rangle = N_{m,\tau}^{(q)}\frac{1}{\sqrt{d}}\sum_{j=0}^{d-1} e^{i\chi_{m,j}^{(q)}}|jq\rangle \otimes \begin{pmatrix} 1 \\ \beta_{m,\tau}^{(q)} e^{i2q\varepsilon j} \end{pmatrix},
\end{equation} 
\begin{equation}\label{ev2}
|\psi_{m,\tau}^{(q= 0)}\rangle = N_{m,\tau}^{(0)}~|m\rangle \otimes \begin{pmatrix} 1 \\ \beta_{m,\tau}^{(0)}  \end{pmatrix},
\end{equation} 
where
\begin{equation}
    \beta_{m\tau}^{(q\neq0)}=\frac{(-1)^qe^{i\tau \xi_{m}^{(q)}}e^{-i(m+q)\varepsilon}-\cos\theta e^{-2i(m+q)\varepsilon}e^{i\gamma}}{\sin\theta e^{i\sigma}}
\end{equation}
\begin{equation}
\beta_{m,\tau}^{(q= 0)} =  \frac{e^{i\tau \xi_{m}^{(0)}}-\cos\theta e^{i\sigma}}{\sin\theta e^{i\sigma}},
\end{equation}
\begin{equation}
    \chi_{m,j}^{(q)}=-\frac{q\varepsilon}{2}j^2+(m\varepsilon+q\pi)j,
\end{equation}
\begin{equation}\label{N}
    N_{m,\tau}^{(q)}=\frac{1}{\sqrt{1+|\beta_{m,\tau}^{(q)}|^2}},
\end{equation}
and the vectors $|jq\rangle$ and $|m\rangle$ in Eqs. (\ref{ev1}) and (\ref{ev2}) are given by the following formula
\begin{equation}\label{k-state}
    |k\rangle=\sum_{x=0}^{d-1}e^{ik\varepsilon x}|x\rangle,
\end{equation}
where $k=0,1,\ldots,d-1$.

\begin{figure}[t]
\includegraphics[width=8.5cm]{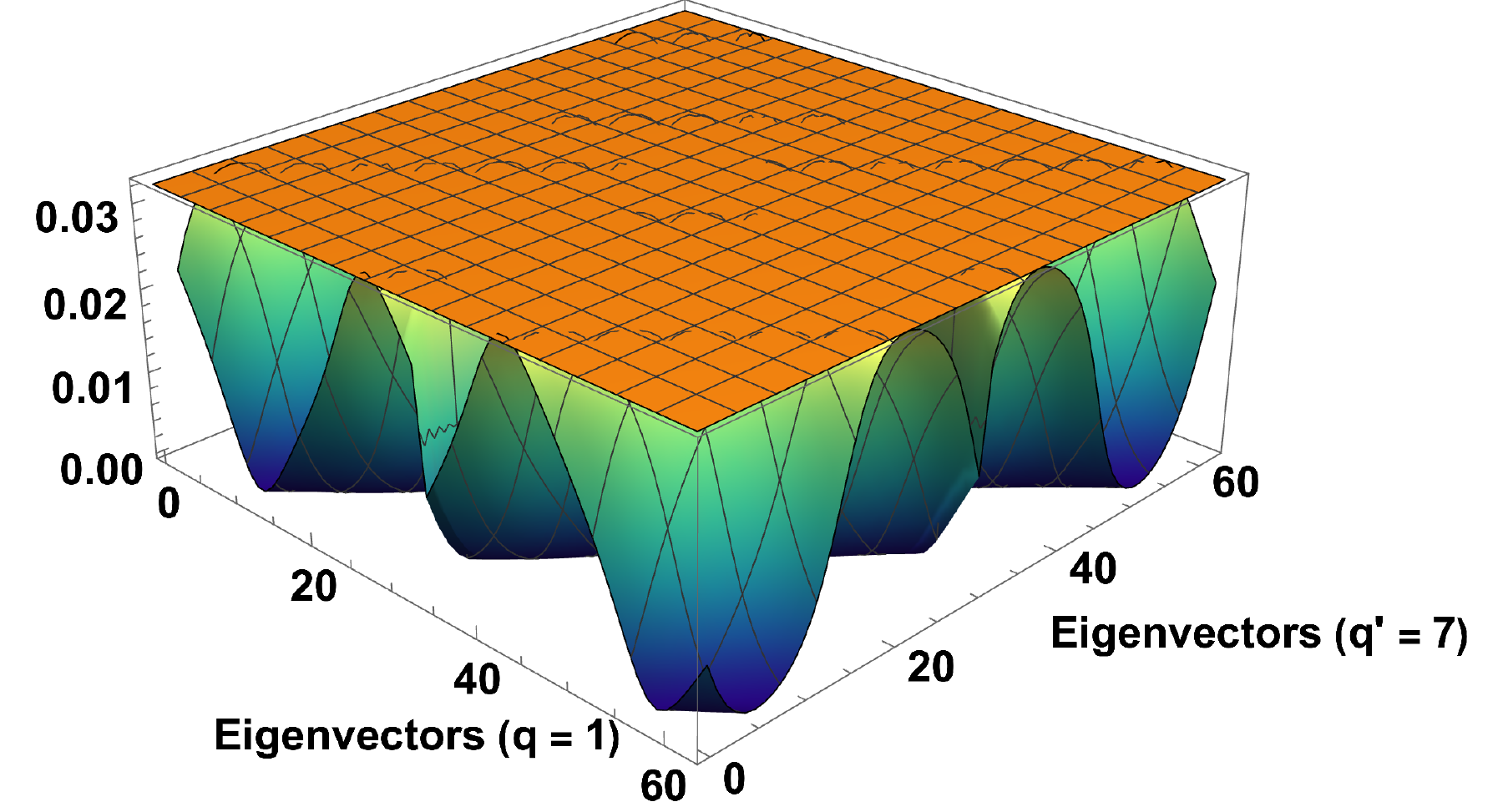}
\includegraphics[width=8.5cm]{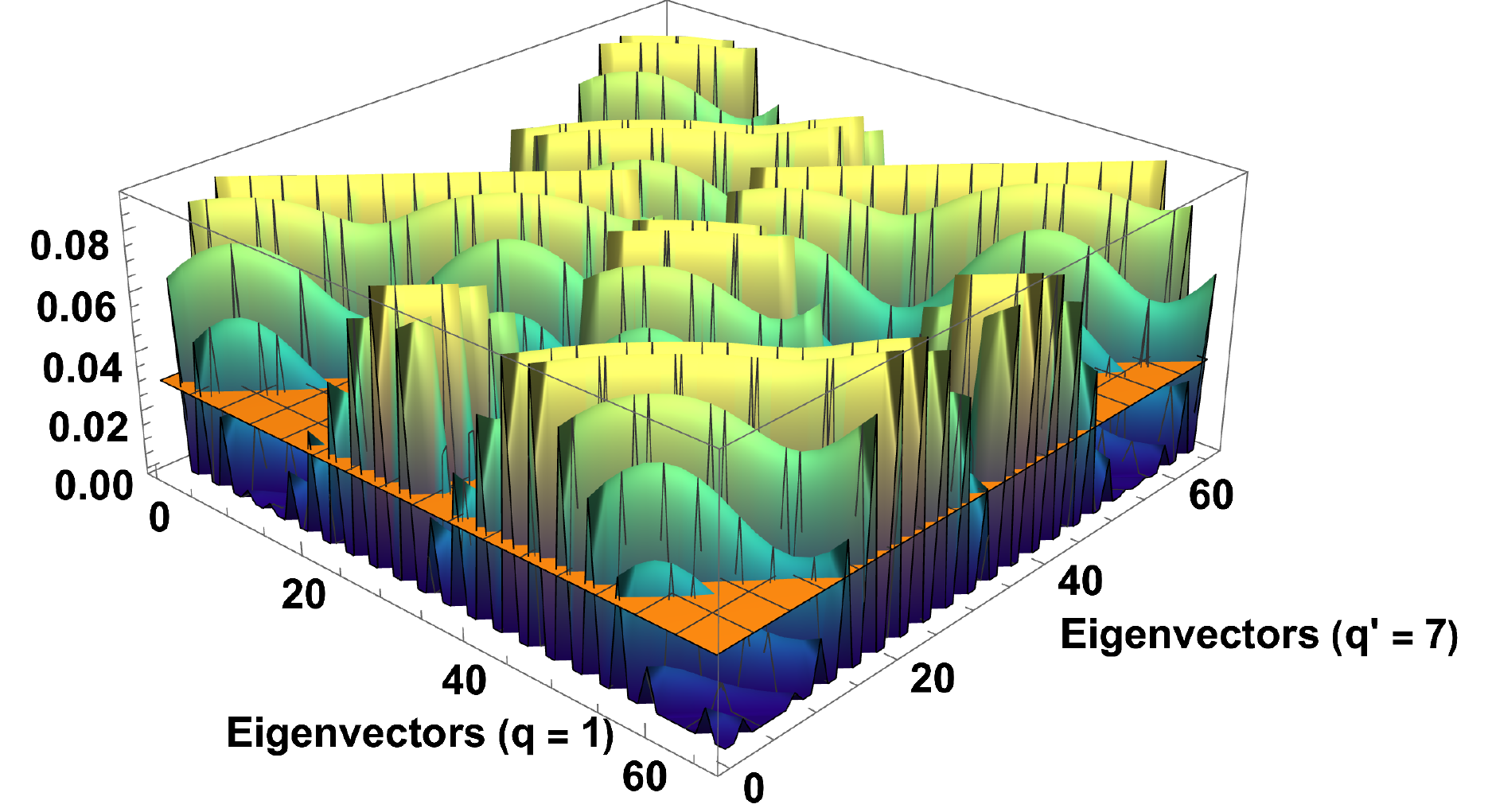}
\caption{The square of the overlap between the eigenvectors $ | \langle\psi_{m',\tau'}^{(q')} |\psi_{m,\tau}^{(q)}\rangle|^2 $ corresponding to $q=1$ and to $q'=7$. The coin tossing operator corresponds to $\gamma=\delta=\sigma=0$ and $\theta = \pi/4$ (see Eq. (\ref{coin})). The size of the cycle is $d=31$ (top) and $d=33$ (bottom). The value of $2/D = 1/d$ is represented by the orange plane. The plots represent a discrete set of values, however the points corresponding to squared overlaps were joined to provide a better visualisation}
 \label{fig1}
\end{figure}

The above results allow us to formulate the following theorem
\begin{theorem}\label{th1}
If $d$ is prime and $q\neq q'$, then for all $m,m',\tau,\tau'$ the following holds
\begin{equation}
 | \langle\psi_{m',\tau'}^{(q')} |\psi_{m,\tau}^{(q)}\rangle| \leq \sqrt{\frac{2}{D}} = \frac{1}{\sqrt{d}}.
\end{equation}
\end{theorem}
This theorem is proven in the Appendix B. In simple words, it states that for prime $d$ the overlap between the eigenvectors corresponding to $q\neq q'$ is never greater than $\sqrt{2/D}$. This implies strong complementarity between the two eigenbases. Nevertheless, not all overlaps are the same, hence the eigenvectors corresponding to $q\neq q'$ are not MUBs. However, the modulus square of their overlap is bounded by twice the inverse of the system's dimension, therefore in large Hlibert spaces the corresponding eigenbases are {\it almost} MUBs (aMUBs). This is clearly visible in an example in Fig. \ref{fig1} (top).  It is natural to ask what happens for non-prime $d$. We observed that for some choices of $q\neq q'$ the overlaps between the corresponding bases are bounded by $\sqrt{2/D}$, however in general this overlap exceeds $\sqrt{2/D}$ -- see Fig.  \ref{fig1} (bottom). Moreover, for non-prime $d$ the overlap between the eigenvectors corresponding to $q\neq q'$ can be quite large. For example, for $\gamma=\delta=\sigma=0$ and $\theta = \pi/4$ it can reach $\sqrt{1/3}$ ($d=18$), or $\sqrt{1/2}$ ($d=16$).

\begin{figure*}
\includegraphics[width=5.5cm]{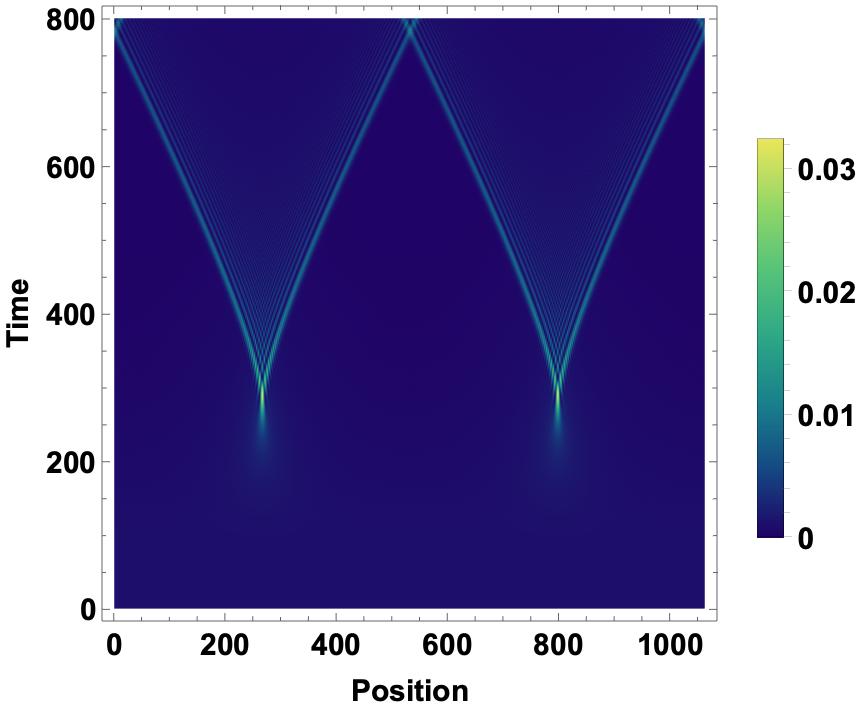} ~~~\includegraphics[width=5.8cm]{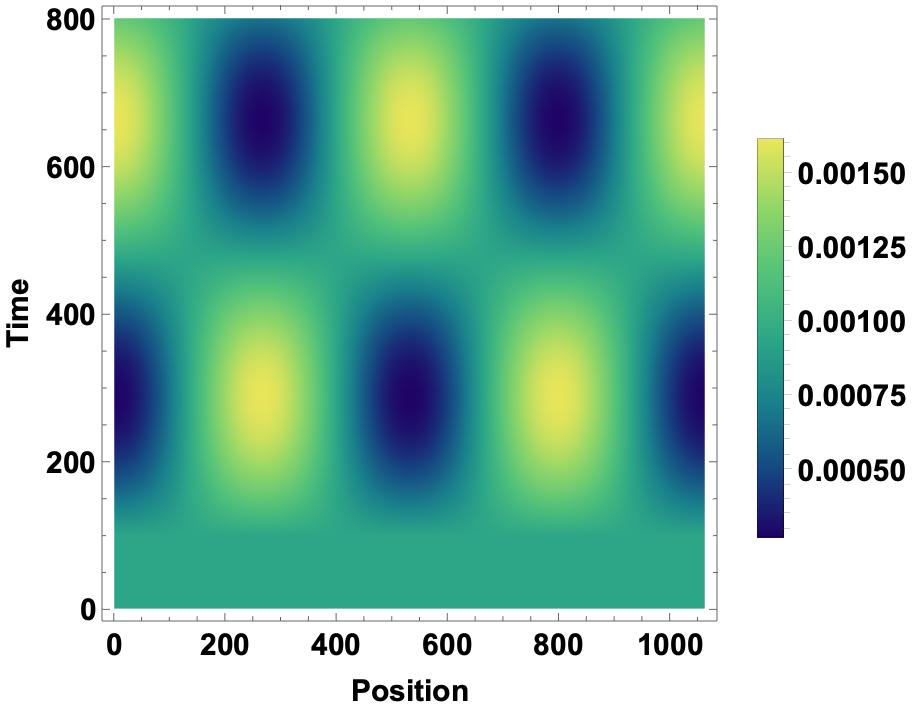} ~~~\includegraphics[width=5.6cm]{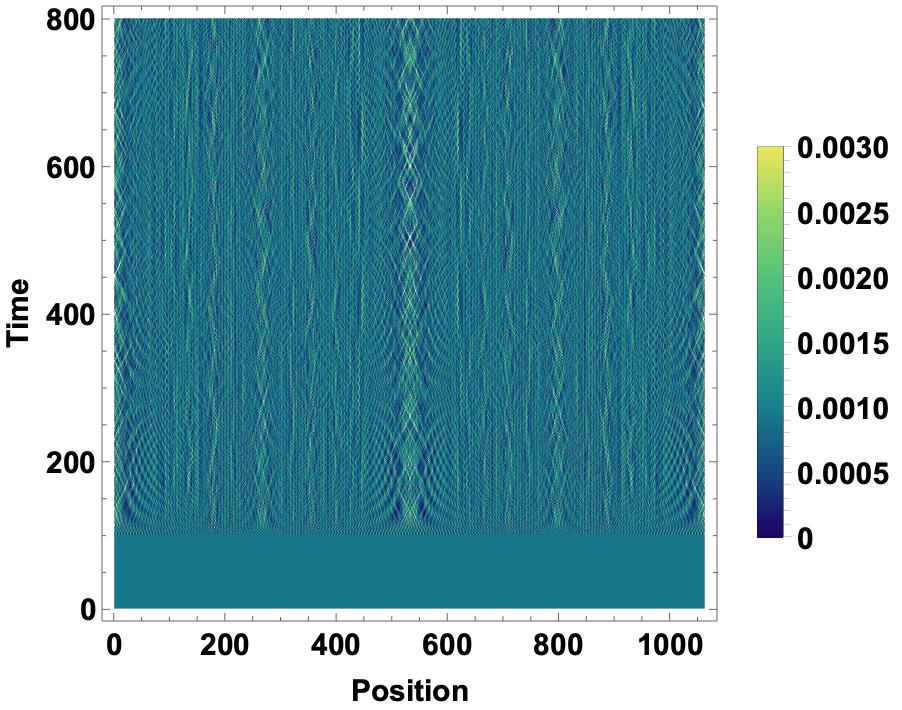} 
\caption{Examples of spatial probability distributions for $800$ steps of dynamics on $d$-cycle ($d=1063$). The coin operator corresponds to $\gamma=\delta=\sigma=0$ and $\theta = \pi/4$ (see Eq. (\ref{coin})). The initial state in all examples is uniformly distributed over all positions. More precisely, it is of the form $|\psi_0\rangle = \frac{1}{\sqrt{2d}}|k=0\rangle \otimes (|+\rangle +i|-\rangle)$ (see Eq. (\ref{k-state})), which is an eigenstate of $U_{q=0}$. In addition, in all examples the first $99$ steps are governed by $U_{q=0}$. Left: the steps $100 \leq t \leq 800$ are governed by $U_{q=1}$. Middle: the step $t=100$ is governed by $U_{q=1}$ and then the steps $101\leq t\leq 800$ are governed by $U_{q=0}$. Right: the steps $100 \leq t \leq 800$ are governed by $U_{q=t}$, i.e., the last 701 steps of the evolution are generated by $U_{q=800}U_{q=799}\ldots U_{q=101} U_{q=100}$.}
 \label{fig2}
\end{figure*}

There is an interesting case corresponding to $\theta = 0$. In this situation the coin operator $\openone \otimes C$ commutes with both, the phase-shift operator $F$ and the conditional translation operator $S$, and its action can be effectively ignored. The eigen-problem corresponding to this case is studied in the Appendix C. In this case the eigenvectors of the evolution operator are of the form 
\begin{equation}
    |\Psi_{m\tau}^{(q)}\rangle = \frac{1}{\sqrt{d}}\sum_{j=0}^{d-1}e^{i\chi_{m\tau j}^{(q)}}|jq\rangle\otimes |\tau\rangle,
\end{equation}
where
\begin{equation}
    \chi_{m\tau j}^{(q)}=-\tau\frac{q\varepsilon}{2}j^2+\left(m\varepsilon+q\pi\right)j,
\end{equation}
$m=0,1,\ldots,d-1$ and $\tau = \pm 1$. The spatial part of these eigenvectors recovers perfect MUB relations, since for $q\neq q'$
\begin{equation}
    |\langle\Psi_{m\tau}^{(q)}|\Psi_{m'\tau'}^{(q')}\rangle|^2 = \delta_{\tau \tau'}\frac{1}{d}.
\end{equation}

The reason for the above stems from the following fact. Note that for $\theta = 0$ we ignore the action of $C$, hence the evolution operator can be written as $U = SF$. In addition, one can represent
\begin{equation}
S = X \otimes \frac{(\openone+\sigma_z)}{2} + X^{\dagger} \otimes \frac{(\openone-\sigma_z)}{2}, 
\end{equation}
and
\begin{equation}
F = Z^{q} \otimes \frac{(\openone+\sigma_z)}{2} + (Z^q)^{\dagger} \otimes \frac{(\openone-\sigma_z)}{2}, 
\end{equation}
where $\sigma_z$ is the Pauli $z$-matrix and 
\begin{eqnarray}
& &X|x\rangle = |x+1 ~\text{mod}~ d\rangle, \\
& &Z|x\rangle = e^{i\frac{2\pi x}{d}}|x\rangle,
\end{eqnarray}
are the Weyl-Heisenberg operators. It is known that for prime $d$ the eigenbases of the following set of Weyl-Heisenberg operators form $d+1$ MUB \cite{DURT:2010wm}
\begin{equation}
\{Z,X,XZ,XZ^2,\ldots,XZ^q,\ldots,XZ^{d-1}\}.
\end{equation}

An immediate dynamical consequence of Theorem 1 is as follows. Let us assume that the system is prepared as some eigenvector of the evolution operator $U_q$, where the index $q$ denotes the corresponding phase-shift's magnitude. Next, assume that after the above preparation the system's dynamics is governed by $U_{q'}$ ($q\neq q'$). The strong complementarity between $U_q$ and $U_{q'}$ implies that the initial state is a superposition over many eigenvectors of $U_{q'}$ and hence the observed dynamics should be far from stationary. Such an example is presented in Fig. \ref{fig2} (left). 

Perhaps an even more striking consequence of Theorem 1 corresponds to the following situation. If the system's evolution is governed by $U_q$ and its initial state is an eigenvector of $U_q$, then nothing happens -- the system is in a stationary state. However, if we suddenly change the evolution operator to $U_{q'}$  the systems's state will be transformed into a superposition of a large number of eigenvectors of $U_q$. Therefore, if after a single application of $U_{q'}$ we come back to the evolution governed by $U_q$, we should observe a sudden onset of dynamics. Moreover, due to strong complementarity between $U_q$ and $U_q'$, this dynamics should significantly depart from the previous stationary state. Such an example is presented in Fig. \ref{fig2} (middle).   

Finally, if in each step we use a different evolution operator $U_q$, for example at time $t$ we apply $U_{q=t}$, then the observed dynamics should exhibit an even more complicated features. This is because of the mutual complementarity between all operators used during the whole evolution. The complementarity occurs not only between eigenvectors of the subsequent operators, but also between the eigenvectors of any two operators applied at two arbitrary moments. This situation is presented in Fig. \ref{fig2} (right). In this example one starts with a uniform distribution and observes an emergence of a nontrivial pattern, the origin of which remains to be explained. 

We should also mention that the strong complementarity implied by Theorem 1 is a sufficient, not a necessary condition for the above behaviour. It is enough that the eigenvector playing the role of the initial state has small overlaps with eigenvectors of unitary operators used in subsequent steps. Theorem 1 guarantees that this happens for every initial eigenvector.  


\section*{Dirac particle analogy}

The above DTQW exhibits the strong complementarity property if its dimension is a doubled prime. One may ask if this finding is just a peculiarity, resulting from a discrete spacetime formulation, that might disappear in the continuous limit. To answer this question we considered the continuous limit for a particular coin operator, corresponding to  $\gamma=\delta=\sigma=0$, which was shown to describe a Dirac particle \cite{KURZYNSKI20086125,PhysRevA.73.054302} . In this case the elements of the evolution operator can be rewritten with the help of position, momentum and Pauli operators as
\begin{equation}
S = e^{i p \sigma_z},~~C=-ie^{i \frac{\pi}{2}(s \sigma_x + c \sigma_z)},~~F=e^{i \varphi x \sigma_z},
\end{equation}
and that the above DTQW evolution operator can emerge as a result of a Trotterisation of 
\begin{equation}\label{Dirac1}
U = e^{-i H \Delta t} = e^{-i \left((p-e A) \sigma_z + m \sigma_x\right) \Delta t},
\end{equation}
where $-\Delta t = 1$, $eA\Delta t = \varphi x + \frac{\pi}{2}c$, $-m \Delta t = \frac{\pi}{2}s$, and we omitted the global phase of $-i$.  The above Hamiltonian $H$ is an analogue of the one-dimensional Dirac Hamiltonian, for which $e$ is the particle's charge, $m$ the mass, and $A$ is the $x$-component of the vector potential. 

We introduce $A = \mu x$, where $\mu$ is a continuous parameter that is an analog of the discrete $\varphi$.  The evolution operator (\ref{Dirac1}) is associated to the following Dirac equation
\begin{equation}\label{Dirac2}
\begin{pmatrix} -i\frac{\partial }{\partial x} - \mu x & m \\ m & i\frac{\partial }{\partial x} + \mu x  \end{pmatrix} \begin{pmatrix} \alpha(x) \\ \beta(x) \end{pmatrix} = E \begin{pmatrix} \alpha(x) \\ \beta(x) \end{pmatrix},
\end{equation}
where $\alpha(x)$ and $\beta(x)$ are the two spinor components of the eigenfunction and $E$ is the system's energy. We adopted the standard convention in which $\hbar = 1$ and the velocity of light $c=1$. We also assumed that the particle's charge is $e=1$. 

The solution of (\ref{Dirac2}) is provided in the Appendix D. The energies of the particle are given by
\begin{equation}\label{DE}
E_{k,\pm} = \pm \sqrt{k^2 + m^2},
\end{equation}
where $-\infty <k< \infty$ is the particle's momentum. On the other hand, the corresponding eigenfunctions are
\begin{equation}\label{DF}
\psi_{k,\pm}^{\mu}(x) = {\mathcal N}_{k,\pm} e^{i(\frac{\mu}{2} x^2 + kx)}\begin{pmatrix} m \\ \pm \sqrt{k^2+m^2}-k \end{pmatrix},
\end{equation}
where
\begin{equation}\label{DN}
{\mathcal N}_{k,\pm} = 2(k^2+m^2 \mp k\sqrt{k^2+m^2}).
\end{equation}
Interestingly, only the eigenfunctions depend on $\mu$. Since the particle is unbounded ($-\infty <x< \infty$), the above eigenfunctions are unnormalised and we have $|\psi^{\mu}_{k,\pm}(x)|^2=1$, but $\int_{-\infty}^{\infty} dx|\psi_{k,\pm}^{\mu}(x)|^2=\infty$.
 
Now we are able to formulate the next theorem
\begin{theorem}\label{th2}
For all $\mu\neq \mu',k,k'$ and $z,z'=\pm 1$ the following holds
\begin{equation}
\left| \int_{-\infty}^{\infty} dx \psi^{\mu}_{k,z}(x)\left(\psi^{\mu'}_{k',z'}(x)\right)^{\ast} \right| \leq \sqrt{\frac{2\pi}{|\mu - \mu'|}}.
\end{equation}
\end{theorem} 
This theorem states that the overlap between any two eigenvectors corresponding to two different Dirac Hamiltonians (with $\mu \neq \mu'$) never exceeds a certain finite value. It is a continuous analog of Theorem 1. Its proof is given in Appendix E.
 
At this point it is worth to relate the above result to a study of MUBs in continuous-variable systems \cite{PhysRevA.78.020303}. It is known that the eigenfunctions of the position and momentum operators are complementary and that in fact the position and momentum eigenbases are MUBs since 
\begin{equation}
\langle x|p\rangle = \frac{1}{\sqrt{2\pi}}e^{i x p}.
\end{equation}
Once again we assume that $\hbar = 1$. It was shown in \cite{PhysRevA.78.020303} that if one defines the operator
\begin{equation}
x_{\theta} = x \cos{\theta} + p \sin{\theta} ,
\end{equation}
then all eigenvectors of $x_{\theta}$ and $x_{\theta'}$ obey
\begin{equation}\label{relation}
\left| \langle x_{\theta}|x_{\theta'}\rangle \right| = \frac{1}{\sqrt{2\pi |\sin(\theta - \theta')|}}.
\end{equation}
It was concluded that the above bases are not exactly MUBs, since the overlap depends on the difference $\theta-\theta'$. Nevertheless, these bases exhibit many properties of MUBs in a sense that the overlaps between any two vectors from two different bases are the same and are finite, despite the fact that the vectors are unnormalised.

In the Dirac particle case the situation is similar, although the overlaps are not the same. That is why we believe that it is justified to call these bases {\it almost} MUBs (aMUBs), just as we called them in the DTQW case. In addition, note that if we substitute $\gamma \sin\theta = 1$ and $\gamma \cos\theta = -\mu$, so that $\gamma = 1/\sin\theta$ and $\theta = \cot^{-1}(-\mu)$, then our Dirac Hamiltonian can be rewritten as
\begin{equation}
H = \gamma x_{\theta} \sigma_z + m \sigma_x.
\end{equation}
The above clearly shows that in our case the complementarity properties of $x_{\theta}$ and $x_{\theta'}$ are affected by the addition of a two-dimensional spinor space, which results in the imperfect aMUB relation stated in Theorem 2. However, if we consider massless particles ($m=0$) the Dirac Hamiltonian becomes
\begin{equation}
H = \gamma x_{\theta} \sigma_z
\end{equation}
and the relations (\ref{relation}) are recovered. This is in perfect analogy to the DTQW scenario with $\theta = 0$ studied in Appendix C.


\section*{Outlook}

We have shown that a strong complementarity exists in DTQWs and in relativistic systems described by one-dimensional Dirac Hamiltonians. The observation of various dynamical behaviours partially represented in Fig. \ref{fig2} suggests that this complementarity can be related to quantum complex behaviour, such as thermalisation or ergodicity breaking. If this is the case, it would be interesting to find out how much complementarity is needed to observe the onset of such behaviors.


\section*{Acknowledgements}

This research is supported by the Polish National Science Centre (NCN) under the Maestro Grant no. DEC-2019/34/A/ST2/00081.  J.W. acknowledges support from IDUB BestStudentGRANT (NO. 010/39/UAM/0010). Part of numerical studies in this work have been carried out using resources provided by Wroclaw Centre for Networking and Supercomputing (wcss.pl), Grant No. 551 (A.S.S.).


\section*{Appendix A}

Here we derive the eigenvalues and the eigenvectors of the unitary evolution operator that governs the dynamics of out model. Recall that we consider the evolution operator of the form
\begin{equation}
    U = S (I\otimes C) F,
\end{equation}
with the phase operator
\begin{equation}
    F = e^{i\varphi x \sigma_z} = \sum_x |x\rangle\langle x| \otimes \begin{pmatrix}
    e^{i\varphi x} & 0 \\ 0 & e^{-i\varphi x} \end{pmatrix},
\end{equation}
where
\begin{equation}
    \varphi = 2\pi\frac{q}{p}=nq\varepsilon,~~~~\varepsilon = \frac{2\pi}{d},~~~~np=d.
\end{equation}
Step operator is given by
\begin{equation}
    S |x,\rightarrow \rangle = |x+1,\rightarrow \rangle,
\end{equation}
\begin{equation}
    S |x,\leftarrow \rangle = |x-1,\leftarrow \rangle.
\end{equation}
We use the following notation
\begin{equation}
    a_x |x,\rightarrow \rangle + b_x |x,\leftarrow \rangle = |x\rangle \otimes \begin{pmatrix} a_x\\b_x\end{pmatrix}.
\end{equation}
The coin operator is
\begin{equation}
    C = e^{i\delta} \begin{pmatrix}c & s\\-s^{*} & c^{*}\end{pmatrix}\in \mathcal{U}(2),
\end{equation}
where
\begin{equation}
    c=\cos\theta e^{i\gamma},~~~~s=\sin\theta e^{i\sigma},~~~~\det C = e^{2i\delta},
\end{equation}
and we chose
\begin{equation}
    0\leq\theta\leq\frac{\pi}{2}.
\end{equation}
In the following we will consider only coins from the $\mathcal{SU}(2)$
\begin{equation}
    C =\begin{pmatrix}c & s\\-s^{*} & c^{*}\end{pmatrix},
\end{equation}
because parameter $\delta$ only shifts energies and can be included after diagonalization. We will use the {\it momentum} states
\begin{equation}
    |k\rangle =\frac{1}{\sqrt{d}}\sum_{x=0}^{d-1}e^{ik\varepsilon x}|x\rangle,
\end{equation}
$k=0,1,...,d-1$.\newline\noindent
Note how previously defined operators act on these states
\begin{equation}
    S|k\rangle|\rightarrow\rangle = e^{-ik\varepsilon}|k\rangle|\rightarrow\rangle,
\end{equation}
\begin{equation}
    S|k\rangle|\leftarrow\rangle = e^{ik\varepsilon}|k\rangle|\leftarrow\rangle,
\end{equation}
\begin{equation}
    F|k\rangle|\rightarrow\rangle = |k+nq\rangle|\rightarrow\rangle,
\end{equation}
\begin{equation}
    F|k\rangle|\leftarrow\rangle = |k-nq\rangle|\leftarrow\rangle.
\end{equation}
Now let us make an ansatz on the form of eigenvectors
\begin{equation}
    |\Psi_r\rangle = \sum_{j=0}^{p-1}e^{i\chi_j}|r+jqn\rangle\otimes \begin{pmatrix}1\\ \beta_r e^{i\eta_j}\end{pmatrix},
\end{equation}
where
\begin{equation}
    r=0,1,...,n-1.
\end{equation}
Note that
\begin{equation}
    F|\Psi_r\rangle = \sum_{j=0}^{p-1}e^{i\chi_{j-1}}|r+jqn\rangle\otimes \begin{pmatrix}1\\ \beta_r e^{i(-\chi_{j-1}+\chi_{j+1}+\eta_{j+1})}\end{pmatrix}.
\end{equation}
It follows that
\begin{eqnarray}
    & & U|\Phi_r\rangle=\sum_{j=0}^{p-1}e^{i\chi_{j-1}}|r+jqn\rangle \otimes \begin{pmatrix}e^{-i(r+jqn)\varepsilon}&0\\0&e^{i(r+jqn)\varepsilon}\end{pmatrix} \nonumber \\ 
    & & \times  \begin{pmatrix}c & s\\-s^{*} & c^{*}\end{pmatrix}\begin{pmatrix}1\\ \beta_r e^{i(-\chi_{j-1}+\chi_{j+1}+\eta_{j+1})}\end{pmatrix},
\end{eqnarray}
which should be equal to 
\begin{equation}
    \lambda |\Phi_r\rangle=\lambda \sum_{j=0}^{p-1}e^{i\chi_j}|r+jqn\rangle\otimes \begin{pmatrix}1\\ \beta_r e^{i\eta_j}\end{pmatrix}.
\end{equation}
Therefore, our goal is to solve the equation 
\begin{eqnarray}
    & & e^{i\chi_{j-1}} \begin{pmatrix}e^{-i(r+jqn)\varepsilon}&0\\0&e^{i(r+jqn)\varepsilon}\end{pmatrix}   \begin{pmatrix}c & s\\-s^{*} & c^{*}\end{pmatrix} \\
    & &\times \begin{pmatrix}1\\ \beta_r e^{i(-\chi_{j-1}+\chi_{j+1}+\eta_{j+1})}\end{pmatrix}=\lambda e^{i\chi_j}\otimes \begin{pmatrix}1\\ \beta_r e^{i\eta_j}\end{pmatrix}, \nonumber 
\end{eqnarray}
which can be rearranged in the form
\begin{eqnarray}
  & &  \begin{pmatrix}c & s\\-s^{*} & c^{*}\end{pmatrix}\begin{pmatrix}1\\ \beta_r e^{i(-\chi_{j-1}+\chi_{j+1}+\eta_{j+1})}\end{pmatrix}=  \\
    & &\lambda e^{i\chi_j}e^{-i\chi_{j-1}}\begin{pmatrix}e^{i(r+jqn)\varepsilon}&0\\0&e^{-i(r+jqn)\varepsilon}\end{pmatrix}\begin{pmatrix}1\\ \beta_r e^{i\eta_j}\end{pmatrix}. \nonumber
\end{eqnarray}
The above is the system of two equations
\begin{eqnarray}
    & & c+s\beta_r e^{i(-\chi_{j-1}+\chi_{j+1}+\eta_{j+1})} = \nonumber \\ 
    & & \lambda e^{i\chi_j}e^{-i\chi_{j-1}} e^{ir\varepsilon}e^{ijqn\varepsilon},
\end{eqnarray}
and
\begin{eqnarray}
    & & -s^{*}+c^{*}\beta_r e^{i(-\chi_{j-1}+\chi_{j+1}+\eta_{j+1})} = \nonumber \\ 
    & & \lambda e^{i\chi_j}e^{-i\chi_{j-1}} e^{i\eta_j} e^{-ir\varepsilon}e^{-ijqn\varepsilon}\beta_r.
\end{eqnarray}
We insist that exponents should be $j$ independent. This gives us three equations with constants $\mathcal{A}, \mathcal{B}, \mathcal{C}$
\begin{equation}\label{a}
    \chi_{j-1}-\chi_j-jnq\varepsilon=\mathcal{A},
\end{equation}
\begin{equation}\label{b}
    \chi_{j+1}+\eta_{j+1}-\chi_{j-1}=\mathcal{B},
\end{equation}
\begin{equation}\label{c}
    \chi_{j-1}-\chi_j-\eta_{j}+jnq\varepsilon=\mathcal{C}.
\end{equation}
Now our system of equations reads
\begin{equation}
    c+s\beta_r e^{i\mathcal{B}} = \lambda e^{-i\mathcal{A}} e^{ir\varepsilon},
\end{equation}
\begin{equation}
    -s^{*}+c^{*}\beta_r e^{i\mathcal{B}} = \lambda e^{-i\mathcal{C}} e^{-ir\varepsilon}\beta_r.
\end{equation}
Therefore
\begin{equation}
    \beta_r=\frac{\lambda e^{-i\mathcal{A}}e^{-ir\varepsilon}-c}{s}e^{-i\mathcal{B}},
\end{equation}
and we obtain quadratic formula for eigenvalues $\lambda$
\begin{equation}\label{quad}
    \lambda^2-\lambda(ce^{i\mathcal{A}}e^{-ir\varepsilon}+c^{*}e^{i(\mathcal{B}+\mathcal{C})}e^{ir\varepsilon})+e^{i(\mathcal{A}+\mathcal{B}+\mathcal{C})}=0.
\end{equation}
Now we make another ansatz on the form of $\chi_j$ and $\eta_j$
\begin{equation}\label{chi}
    \chi_j=Aj^2+Bj,
\end{equation}
\begin{equation}\label{eta}
    \eta_j=Cj,
\end{equation}
with $A, B, C$ cosntant. Plugging Eqs.(\ref{chi},\ref{eta}) to Eq.(\ref{a}) one obtains
\begin{equation}
    -(2A+nq\varepsilon)j+(A-B)=\mathcal{A},
\end{equation}
hence
\begin{equation}
    A=-\frac{nq\varepsilon}{2},
\end{equation}
\begin{equation}
    A-B=\mathcal{A}.
\end{equation}
On the other hand plugging Eqs.(\ref{chi},\ref{eta}) to Eq.(\ref{b}) leads to
\begin{equation}
    (4A+C)j+(2B+C)=\mathcal{B},
\end{equation}
hence
\begin{equation}
    C=2nq\varepsilon.
\end{equation}
Again plugging Eqs.(\ref{chi},\ref{eta}) to Eq.(\ref{c}) gives us following formula
\begin{equation}
    -(2A+C-nq\varepsilon)j+(A-B)=\mathcal{C},
\end{equation}
therefore
\begin{equation}
    \mathcal{C}=A-B=\mathcal{A}.
\end{equation}
We demand that periodic condition should be fulfilled i.e. $\chi_{j+p}-\chi_j$ must be integer multiple of $2\pi$. We have 
\begin{equation}
    \chi_{j+p}-\chi_j= Ap^2+Bp+2 Apj
\end{equation}
and by taking into account that $2 A p j=-jq2\pi$ we get condition
\begin{equation}
    B=\frac{2\pi m}{p}-Ap=mn\varepsilon+q\pi,
\end{equation}
with
\begin{equation}
    m=0,1,...,p-1.
\end{equation}
Let us summarize what we obtained so far 
\begin{equation}
    A=-\frac{1}{2}nq\varepsilon,~~B=mn\varepsilon+q\pi,~~C=2nq\varepsilon,
\end{equation}
\begin{equation}
    \mathcal{C}=\mathcal{A}=\frac{qn\varepsilon}{2}-((m+q)n\varepsilon+q\pi),
\end{equation}
\begin{equation}
    \mathcal{B}=2((m+q)n\varepsilon+q\pi).
\end{equation}
In our quadratic formula Eq.(\ref{quad}) linear coefficient can be now expressed as
\begin{equation}
    ce^{i\mathcal{A}}e^{-ir\varepsilon}+c^{*}e^{i(\mathcal{B}+\mathcal{C})}e^{ir\varepsilon}=2e^{iqn\varepsilon/2}\cos \xi,
\end{equation}
where
\begin{equation}
    \cos \xi=\cos\theta\cos\Omega,
\end{equation}
\begin{equation}
    \Omega=\gamma-\Xi-r\varepsilon,
\end{equation}
\begin{equation}
    \Xi=(m+q)n\varepsilon+q\pi.
\end{equation}
Now the formula for eigenvalues take the form
\begin{equation}
    \lambda^2-2\lambda e^{iqn\varepsilon/2}\cos\xi+e^{iqn\varepsilon},
\end{equation}
which can be easly solved
\begin{equation}
    \lambda_\pm=e^{iqn\varepsilon/2}e^{\pm i \xi}.
\end{equation}
We also have
\begin{equation}
    \beta_r=\frac{\lambda e^{-i\left(\frac{qn\varepsilon}{2}+\Xi\right)}e^{ir\varepsilon}-e^{-2i\Xi}c}{s}.
\end{equation}


\section*{Appendix B}

In this appendix we provide a proof of the Theorem 1. 

\subsubsection*{Case $q'\neq0$, $q\neq0$.}

We first observe that
\begin{eqnarray}
  & &  \langle\psi_{m',\tau'}^{(q')} |\psi_{m,\tau}^{(q)}\rangle =N_{m',\tau'}^{(q')}N_{m,\tau}^{(q)}\frac{1}{d}\sum_{j,j'=0}^{d-1} e^{i(\chi_{m,j}^{(q)}-\chi_{m',j'}^{(q')})} \nonumber \\ 
    & &\times \langle j'q'|jq\rangle
    \begin{pmatrix} 1 \\ \beta_{m',\tau'}^{(q')} e^{i2q'\varepsilon j'} \end{pmatrix}^\dag \begin{pmatrix} 1 \\ \beta_{m,\tau}^{(q)} e^{i2q\varepsilon j} \end{pmatrix}.
\end{eqnarray}
For a given pair $q,q'$ ($q',q<d~~$, $~~q'\neq q$) and $j$ let us define unique $0\leq \Tilde{j}\leq d$ such that $\Tilde{j}q'$ is congruent with $jq$ modulo $d$
\begin{equation}\label{j}
    \Tilde{j}q'\equiv jq~ (mod~d).
\end{equation}
Note that
\begin{equation}
    \langle j'q'|jq\rangle=\delta_{j'\Tilde{j}},
\end{equation}
where we use Kronecker delta $\delta_{j'\Tilde{j}}$. Now
\begin{eqnarray}
& &  \langle\psi_{m',\tau'}^{(q')} |\psi_{m,\tau}^{(q)}\rangle =N_{m',\tau'}^{(q')}N_{m,\tau}^{(q)}\frac{1}{d}\sum_{j=0}^{d-1} e^{i(\chi_{m,j}^{(q)}-\chi_{m',\Tilde{j}}^{(q')})} \nonumber \\
& & \times  \begin{pmatrix} 1 \\ \beta_{m',\tau'}^{(q')} e^{i2q'\varepsilon \Tilde{j}} \end{pmatrix}^\dag \begin{pmatrix} 1 \\ \beta_{m,\tau}^{(q)} e^{i2q\varepsilon j} \end{pmatrix}.
\end{eqnarray}
Eq. (\ref{j}) implies
\begin{equation}
    e^{i2\varepsilon (qj-q'\Tilde{j})}=1,
\end{equation}
therefore
\begin{equation}
    \begin{pmatrix} 1 \\ \beta_{m',\tau'}^{(q')} e^{i2q'\varepsilon \Tilde{j}} \end{pmatrix}^\dag \begin{pmatrix} 1 \\ \beta_{m,\tau}^{(q)} e^{i2q\varepsilon j} \end{pmatrix}=
    1+(\beta_{m',\tau'}^{(q')})^*\beta_{m,\tau}^{(q)}.
\end{equation}
We arrive at
\begin{eqnarray}
 & &   \langle\psi_{m',\tau'}^{(q')} |\psi_{m,\tau}^{(q)}\rangle =   
    N_{m',\tau'}^{(q')}N_{m,\tau}^{(q)}\left(1+(\beta_{m',\tau'}^{(q')})^*\beta_{m,\tau}^{(q)}\right) \nonumber \\
    & & \times \frac{1}{d}\sum_{j=0}^{d-1} e^{i(\chi_{m,j}^{(q)}-\chi_{m',\Tilde{j}}^{(q')})}
\end{eqnarray}
and
\begin{eqnarray}
 & &   | \langle\psi_{m',\tau'}^{(q')} |\psi_{m,\tau}^{(q)}\rangle|^2=\frac{1}{d^2}~\left(N_{m',\tau'}^{(q')}N_{m,\tau}^{(q)}\right)^2 \nonumber \\
  & &  \times \left|1+(\beta_{m',\tau'}^{(q')})^*\beta_{m,\tau}^{(q)}\right|^2\left|\sum_{j=0}^{d-1} e^{i(\chi_{m,j}^{(q)}-\chi_{m',\Tilde{j}}^{(q')})}\right|^2.
\end{eqnarray}

Next, let us observe that (the proof is given latter)
\begin{equation}\label{S}
\left|\sum_{j=0}^{d-1} e^{i(\chi_{m,j}^{(q)}-\chi_{m',\Tilde{j}}^{(q')})}\right|^2 = d.
\end{equation}
Using the above we get
\begin{equation}
     | \langle\psi_{m',\tau'}^{(q')} |\psi_{m,\tau}^{(q)}\rangle|^2=\frac{1}{d}~\left(N_{m',\tau'}^{(q')}N_{m,\tau}^{(q)}\right)^2~\left|1+(\beta_{m',\tau'}^{(q')})^*\beta_{m,\tau}^{(q)}\right|^2.
\end{equation}
Obviously 
\begin{equation}
\left|1+(\beta_{m',\tau'}^{(q')})^*\beta_{m,\tau}^{(q)}\right|^2\leq \left(1+|\beta_{m',\tau'}^{(q')}||\beta_{m,\tau}^{(q)}|\right)^2,
\end{equation}
therefore (see the proof at the end of this appendix)
\begin{eqnarray}\label{abs}
   & & \left(N_{m',\tau'}^{(q')}N_{m,\tau}^{(q)}\right)^2~\left|1+(\beta_{m',\tau'}^{(q')})^*\beta_{m,\tau}^{(q)}\right|^2 \leq  \nonumber \\ & &\frac{\left(1+|\beta_{m',\tau'}^{(q')}||\beta_{m,\tau}^{(q)}|\right)^2}{\left(1+|\beta_{m',\tau'}^{(q')}|^2\right)\left(1+|\beta_{m,\tau}^{(q)}|^2\right)}\leq 1.
\end{eqnarray}
It follows that
\begin{equation}
    | \langle\psi_{m',\tau'}^{(q')} |\psi_{m,\tau}^{(q)}\rangle|^2\leq \frac{1}{d}=\frac{2}{D}.
\end{equation}

\subsubsection*{Case $q'=0,q\neq 0$}

We have
\begin{eqnarray}
 & & \langle\psi_{m',\tau'}^{(0)} |\psi_{m,\tau}^{(q)}\rangle=N_{m',\tau'}^{(0)}N_{m,\tau}^{(q)}\frac{1}{\sqrt{d}}\sum_{j=0}^{d-1} e^{i\chi_{m,j}^{(q)}} \nonumber \\
 & &\times \langle m'|jq\rangle
    \begin{pmatrix} 1 \\ \beta_{m',\tau'}^{(0)}\end{pmatrix}^\dag \begin{pmatrix} 1 \\ \beta_{m,\tau}^{(q)} e^{i2q\varepsilon j} \end{pmatrix}.
\end{eqnarray}
Since $\langle m'|jq\rangle\neq 0$ for $j=\Tilde{j}$, where $m'\equiv \Tilde{j}q~(mod~d)$
\begin{eqnarray}
  & & \langle\psi_{m',\tau'}^{(0)} |\psi_{m,\tau}^{(q)}\rangle=N_{m',\tau'}^{(0)}N_{m,\tau}^{(q)}\frac{1}{\sqrt{d}}~e^{i\chi_{m,\Tilde{j}}^{(q)}} \nonumber \\ & & \times \left(1+(\beta_{m',\tau'}^{(0)})^*\beta_{m,\tau}^{(q)}e^{i2q\varepsilon \Tilde{j}}\right)
\end{eqnarray}
and
\begin{eqnarray}\label{kw}
& & \left|\langle\psi_{m',\tau'}^{(0)} |\psi_{m,\tau}^{(q)}\rangle\right|^2=\left(N_{m',\tau'}^{(0)}\right)^2\left(N_{m,\tau}^{(q)}\right)^2 \nonumber \\ & & \times \frac{1}{d}~\left|1+(\beta_{m',\tau'}^{(0)})^*\beta_{m,\tau}^{(q)}e^{i2q\varepsilon \Tilde{j}}\right|^2.
\end{eqnarray}
Since
\begin{equation}
    \left|1+(\beta_{m',\tau'}^{(0)})^*\beta_{m,\tau}^{(q)}e^{i2q\varepsilon \Tilde{j}}\right|^2\leq \left(1+|\beta_{m',\tau'}^{(0)}||\beta_{m,\tau}^{(q)}|\right)^2
\end{equation}
one has
\begin{equation}
    \left|\langle\psi_{m',\tau'}^{(0)} |\psi_{m,\tau}^{(q)}\rangle\right|^2\leq\frac{1}{d}~\frac{\left(1+|\beta_{m',\tau'}^{(0)}||\beta_{m,\tau}^{(q)}|\right)^2}{\left(1+|\beta_{m',\tau'}^{(0)}|^2\right)\left(1+|\beta_{m',\tau'}^{(q)}|^2\right)}\leq \frac{1}{d}.
\end{equation}

\subsubsection*{Proof of Eq. (\ref{S})}

\noindent For given $q,q'\neq q$ 
\begin{equation}\label{w}
   S= \sum_{j=0}^{d-1} e^{i(\chi_{m,j}^{(q)}-\chi_{m',\Tilde{j}}^{(q')})}=
    \sum_{j=0}^{d-1} e^{i\varepsilon(w_j+v_{m,j})},
\end{equation}
where 
\begin{equation}
    w_j=\left(\frac{qj(d-j)-q'\Tilde{j}(d-\Tilde{j})}{2}\right)_{mod~d},
\end{equation}

\begin{equation}
    v_{m,j}=(mj-m'\Tilde{j})_{mod~d}.
\end{equation}
Note that $j,\Tilde{j},q,q',m,m'$ can be considered as elements of finite field $\mathbb{F}_d=\mathbb{Z}/ d\mathbb{Z}$. $w_j$ and $v_{m,j}$ are also elements of $\mathbb{F}_d$ (note that $\left[qj(d-j)-q'\Tilde{j}(d-\Tilde{j})\right]$ is even). Let us define another element $Q$ of this field with the equation $Q\circ q'=q$, where $\circ$ stands for the field multiplication. Due to this definition 
\begin{equation}\label{tildj}
    \Tilde{j}=Q\circ j.
\end{equation}
Let us also define one more element of the field $\mathbb{F}_d$, $h$ by $h\circ 2 =1$. Equation (\ref{w}) can be re-framed in the form 
\begin{equation}\label{neww}
    w_j=h\circ\left[(q\circ j\circ(-j))~-_\mathbb{F}~(q'\circ\Tilde{j}\circ(-\Tilde{j}))\right]
\end{equation}
It follows from Eq.(\ref{tildj}) that 
\begin{eqnarray}
    w_j &=& h\circ\left[
    (q'\circ Q \circ Q \circ Q \circ j \circ j-_\mathbb{F}~q\circ j \circ j)
    \right] \nonumber \\ &=& h\circ(Q-_\mathbb{F}~1)\circ q \circ j \circ j=x\circ j \circ j,
\end{eqnarray}
where $x=h\circ(Q-_\mathbb{F}~1)\circ q\in \mathbb{F}_d$. To put it another way $w_j = (xj^2)_{mod~d}$.
Analogously $v_{m,j}=(yj)_{mod~d}$, where $y=m-_\mathbb{F}~m'\circ Q \in \mathbb{F}_d$.

\begin{equation}
    S=\sum_{j=0}^{d-1} e^{i\varepsilon(xj^2+yj)}.
\end{equation}
For $q\neq q'$, $Q\neq 1$ and $x \neq 0$. It follows from theory of generalized quadratic Gauss sums that 
\begin{equation}
    |S|^2=d.
\end{equation}

\subsubsection*{Proof of the second inequality in Eq. (\ref{abs})}

It is clear that
\begin{equation}
    (|a|-|b|)^2\geq0.
\end{equation}
This leads to
\begin{equation}
    |a|^2+|b|^2\geq 2|a||b|,
\end{equation}
and
\begin{equation}
    |a|^2+|b|^2+1+|a|^2|b|^2\geq2|a||b|+1+|a|^2|b|^2,
\end{equation}
which gives
\begin{equation}
     (1+|a|^2)(1+|b|^2) \geq(1+|a||b|))^2.
\end{equation}


\section*{Appendix C}

Now let us solve the eigen-problem for the DTQW with $\theta = 0$. We make an ansatz on the form of eigenvectors
\begin{equation}
    |\Psi_{r,\tau}\rangle = \frac{1}{\sqrt{p}}\left(\sum_{j=0}^{p-1}e^{i\chi_j}|r+jqn\rangle\right)\otimes |\tau\rangle,
\end{equation}
where
\begin{equation}
    r=0,1,...,n-1,~~~\tau = \pm1.
\end{equation}
Note that
\begin{equation}
    F|\Psi_{r,\tau}\rangle = \frac{1}{\sqrt{p}}\left(\sum_{j=0}^{p-1}e^{i\chi_{j-\tau}}|r+jqn\rangle\right)\otimes |\tau\rangle,
\end{equation}
It follows that
\begin{eqnarray}
    U|\Psi_{r,\tau}\rangle &=& \frac{1}{\sqrt{p}}\left(\sum_{j=0}^{p-1}e^{i\chi_{j-\tau}}|r+jqn\rangle\right)  \\
    &\otimes& \begin{pmatrix}e^{-i(r+jqn)\varepsilon}&0\\0&e^{i(r+jqn)\varepsilon}\end{pmatrix} |\tau\rangle \nonumber
\end{eqnarray}
which should be equal to $\lambda  |\Psi_{r,\tau}\rangle$. Therefore, our goal is to solve the equation 
\begin{equation}
    e^{i\chi_{j-\tau}} e^{-i\tau (r+jqn)\varepsilon}=\lambda e^{i\chi_j}.
\end{equation}
We insist that exponents are $j$-independent. This gives us an equation with a constant $\mathcal{A}_\tau$
\begin{equation}\label{a2}
    \chi_{j-\tau}-\chi_j-\tau jnq\varepsilon=\mathcal{A}_\tau,
\end{equation}
and a formula for the eigenvalue $\lambda$
\begin{equation}\label{quad2}
    \lambda = e^{i\mathcal{A}_\tau}e^{-ir\varepsilon\tau}.
\end{equation}
Now we make another ansatz on the form of $\chi_j$
\begin{equation}\label{chi2}
    \chi_j=Aj^2+Bj,
\end{equation}
with $A$, $B$ constant. Plugging Eq.(\ref{chi2}) to Eq.(\ref{a2}) one obtains
\begin{equation}
    -(2A+\tau nq\varepsilon)j+(A-B)=\mathcal{A}_\tau,
\end{equation}
therefore
\begin{equation}
    A=-\frac{\tau nq\varepsilon}{2},
\end{equation}
\begin{equation}
    A-B=\mathcal{A}_\tau.
\end{equation}
We demand that the periodicity condition should be fulfilled, i.e., $\chi_{j+p}-\chi_j$ must be an integer multiple of $2\pi$. We have 
\begin{equation}
    \chi_{j+p}-\chi_j= Ap^2+Bp+2 Apj
\end{equation}
and since $2 A p j=-\tau jq2\pi$ we get
\begin{equation}
    B=\frac{2\pi m}{p}-Ap=mn\varepsilon+\tau q\pi,
\end{equation}
where
\begin{equation}
    m=0,1,...,p-1.
\end{equation}

Let us summarise what we obtained so far:
\begin{equation}
    A=-\tau\frac{nq\varepsilon}{2},~~B=mn\varepsilon+\tau q\pi,
\end{equation}
\begin{equation}
    \mathcal{A}_\tau = A-B = -\tau \frac{nq\varepsilon}{2}-mn\varepsilon-\tau q\pi.
\end{equation}
The eigenvalue formula takes the form
\begin{equation}
    \lambda_{r,m,\tau} = e^{-i(\tau\frac{nq\varepsilon}{2}+mn\varepsilon+q\pi+r\varepsilon)}.
\end{equation}
In addition
\begin{equation}
    \chi_j=-\tau\frac{nq\varepsilon}{2}j^2+\left(mn\varepsilon+q\pi\right)j.
\end{equation}

\subsubsection*{Summary $\theta=0$, $d$ prime}

We get
\begin{equation}
    \lambda_{m,\tau}^{(q)} = e^{-i(\tau\frac{q\varepsilon}{2}+m\varepsilon+q\pi)},
\end{equation}
\begin{equation}
    m=0,...,d-1,
\end{equation}
\begin{equation}
    |\Psi_{m\tau}^{(q)}\rangle = \frac{1}{\sqrt{d}}\sum_{j=0}^{d-1}e^{i\chi_{m\tau j}^{(q)}}|jq\rangle\otimes |\tau\rangle,
\end{equation}
\begin{equation}
    \chi_{m\tau j}^{(q)}=-\tau\frac{q\varepsilon}{2}j^2+\left(m\varepsilon+q\pi\right)j.
\end{equation}
The spatial part of these eigenvectors obeys perfect MUB relations for $q \neq q'$
\begin{equation}\label{MUBrel}
    |\langle\Psi_{m\tau}^{(q)}|\Psi_{m'\tau'}^{(q')}\rangle|^2 = \delta_{\tau \tau'}\frac{1}{d}
\end{equation}

\subsubsection*{Proof of Eq. (\ref{MUBrel})}

Let us write
\begin{equation}
    |\Psi_{m\tau}^{(q)}\rangle = |\Tilde{\Psi}_{m\tau}^{(q)}\rangle \otimes |\tau\rangle,
\end{equation}
where
\begin{equation}
    |\Tilde{\Psi}_{m\tau}^{(q)}\rangle = \frac{1}{\sqrt{d}}\sum_{j=0}^{d-1}e^{i\chi_{m\tau j}^{(q)}}|jq\rangle.
\end{equation}
We need to show that
\begin{equation}
     |\langle \Tilde{\Psi}_{m\tau}^{(q)}|\Tilde{\Psi}_{m'\tau}^{(q')}\rangle|^2 =\frac{1}{d^2}\left|\sum_{j=0}^{d-1}e^{i(\chi_{m\tau j}^{(q)}-\chi_{m'\tau \Tilde{j}}^{(q')})}\right|^2=\frac{1}{d},
\end{equation}
where $\Tilde{j}$ is defined by the equation
\begin{equation}
    \Tilde{j}q' \equiv jq (mod~d).
\end{equation}
However, the above is equivalent to Eq. (\ref{S}), that was already proven in Appendix B.


\section*{Appendix D}

As shown in the manuscript, the continuous limit of our DTQW model leads to the following Dirac equation
\begin{equation}
\begin{pmatrix} -i\frac{\partial }{\partial x} - \mu x & m \\ m & i\frac{\partial }{\partial x} + \mu x  \end{pmatrix} \begin{pmatrix} \alpha \\ \beta \end{pmatrix} = E \begin{pmatrix} \alpha \\ \beta \end{pmatrix},
\end{equation}
where $\alpha = \alpha(x)$, $\beta = \beta(x)$, $m$ is the particle's mass and $\mu$ describes the amplitude of the potential $A = \mu x$. It is assumed that $\hbar = c=1$ and that the particle's charge is $e=1$. The above leads to
\begin{equation}
\beta = \frac{i}{m} \frac{\partial \alpha}{\partial x} + \frac{(\mu x + E)}{m} \alpha,
\end{equation}
and
\begin{equation}\label{DirEq}
\frac{\partial^2 \alpha}{\partial x^2} - 2 i\mu x \frac{\partial \alpha}{\partial x} + (E^2-m^2-\mu^2 x^2 - i\mu)\alpha = 0.
\end{equation}
We make the following ansatz
\begin{equation}
\alpha = {\mathcal N} e^{if(x)},
\end{equation}
where $f(x)$ is some real function and $\mathcal{N}$ is a normalization constant. Since we consider at most a second derivative over $x$, it is enough to take $f(x) = ax^2 + bx +c$ with $a$, $b$ and $c$ being real parameters. We take
\begin{equation}
f(x) = \frac{\mu}{2}x^2 + kx,
\end{equation}
where $k$ is a real parameter, and plug it into Eq. (\ref{DirEq}). We obtain 
\begin{equation}
E_{k,\pm}= \pm \sqrt{k^2 + m^2}.
\end{equation} 
Moreover, the solution is
\begin{equation}
\psi_{k,\pm}^{\mu} = \begin{pmatrix} \alpha_{k,\pm}^{\mu}(x) \\ \beta_{k,\pm}^{\mu}(x) \end{pmatrix},
\end{equation}
where
\begin{equation} 
\alpha_{k,\pm}^{\mu}(x) = {\mathcal N} e^{i\left(\frac{\mu}{2}x^2 + kx \right)}
\end{equation}
and
\begin{equation}
\beta_{k,\pm}^{\mu}(x) = \alpha_{k,\pm}^{\mu}(x) \left(\frac{E-k}{m}\right).
\end{equation}

The equation describes the particle moving in the unbounded space, therefore the solution is not normalized, i.e., 
\begin{equation}
\int_{-\infty}^{+\infty} dx \begin{pmatrix} \alpha^{\ast}(x) & \beta^{\ast}(x) \end{pmatrix} \begin{pmatrix} \alpha(x) \\ \beta(x) \end{pmatrix} = \infty.
\end{equation}
However, we choose $\mathcal N$ such that $|\alpha |^2 + |\beta|^2 = 1$. This leads to
\begin{equation}\label{norm}
{\mathcal N}_{k,\pm} = 2(k^2+m^2 \mp k\sqrt{k^2+m^2}).
\end{equation}


\section*{Appendix E}

Now we prove Theorem 2. First note that
\begin{equation}
(\psi_{k',c'}^{\mu' })^{\ast} \psi_{k,c}^{\mu} = \Gamma_{k,k',c,c'}  e^{i\left(\frac{(\mu-\mu')}{2}x^2 + (k-k')x \right)}
\end{equation}
where $c,c' = \pm 1$ and
\begin{equation}
\Gamma_{k,k',c,c'} = {\mathcal N}_{k',c'}{\mathcal N}_{k,c} \left(m^2 + (E_{k,c}-k)(E_{k',c'}-k')\right).
\end{equation}
Note that Eq. (\ref{norm}) implies
\begin{equation}
|\Gamma_{k,k',c,c'}| \leq 1.
\end{equation}

Next, we evaluate the overlap
\begin{eqnarray}
& &\left| \int_{-\infty}^{+\infty} dx (\psi_{k',c'}^{\mu' })^{\ast} \psi_{k,c}^{\mu} \right|= \nonumber \\
& & \left| \Gamma_{k,k',c,c'}  \int_{-\infty}^{+\infty} dx  e^{i\left(\frac{(\mu-\mu')}{2}x^2 + (k-k')x \right)} \right| \nonumber \leq \\
& &  \left| \int_{-\infty}^{+\infty} dx  e^{i\left(\frac{(\mu-\mu')}{2}x^2 + (k-k')x \right)} \right| 
\end{eqnarray}
We introduce
\begin{equation}
\Delta = \frac{\mu-\mu'}{2},~~~~\delta = k-k',
\end{equation}
and evaluate
\begin{eqnarray}
& &\int_{-\infty}^{+\infty} dx  e^{i\left(\Delta x^2 + \delta x \right)}  = \\
& & e^{-i\frac{\delta^2}{4\Delta}} \int_{-\infty}^{+\infty} dx  e^{i\left(\Delta x + \frac{\delta}{2\sqrt{\Delta}} \right)^2} = e^{-i\frac{\delta^2}{4\Delta}} \frac{(1+i)\sqrt{\pi}}{\sqrt{2\Delta}}. \nonumber
\end{eqnarray}
Therefore
\begin{equation}
\left| \int_{-\infty}^{+\infty} dx (\psi_{k',c'}^{\mu' })^{\ast} \psi_{k,c}^{\mu} \right| \leq \sqrt{\frac{2\pi}{|\mu -\mu'|}}.
\end{equation}



\end{document}